\documentclass[10pt, paper=a4, UKenglish]{article}
\usepackage{amsmath}

\usepackage{graphicx}
\def\Title#1{\begin{center} {\Large #1 } \end{center}}
\def\Author#1{\begin{center}{ \sc #1} \end{center}}
\def\Address#1{\begin{center}{ \it #1} \end{center}}

\newenvironment{Abstract}{\begin{quotation} \begin{center} 
             \large ABSTRACT \end{center}\bigskip 
      \begin{center}\begin{large}}{\end{large}\end{center} \end{quotation}}

\def\Acknowledgements{\bigskip  \bigskip \begin{center} \begin{large}
      \bf ACKNOWLEDGEMENTS \end{large}\end{center}}

\def\beq{\begin{equation}}
\def\eeq#1{\label{#1}\end{equation}}
\def\eeqn{\end{equation}}

\def\beqa{\begin{eqnarray}}
\def\eeqa#1{\label{#1}\end{eqnarray}}
\def\eeqan{\end{eqnarray}}

\let\bar=\overbar

\def\Dslash{\not{\hbox{\kern-4pt $D$}}}
\def\dslash{\not{\hbox{\kern-2pt $\del$}}}

\def\msb{{\bar{\ssstyle M \kern -1pt S}}}

\usepackage{color}
\usepackage{lineno}
\usepackage{subfig}
\usepackage{hyperref}

\def\affiliation{
$^1$Department of Physics, Stanford University, USA\\
$^2$Department of Physics, University of California at Berkeley and Lawerence Berkeley National Laboratory, USA}

\usepackage{fancyhdr}
\definecolor{mygrey}{RGB}{105,105,105}
\begin{document}


\large
\begin{titlepage}

\vfill
\Title{Exploration of different parameter optimization algorithms within the context of ACTS software framework}
\vfill

\Author{ Rocky Bala Garg$^{1}$, Heather Gray$^{2}$, Elyssa Hofgard$^{1}$, Lauren Tompkins$^{1}$}
\Address{\affiliation}
\vfill

\begin{Abstract}
Particle track reconstruction, in which the trajectories of charged particles are determined, is a critical and time consuming component of the full event reconstruction chain. The underlying software is complex and consists of a number of mathematically intense algorithms, each dealing with a particular tracking sub-process.
These algorithms have many input parameters that need to be supplied in advance. However, it is difficult to determine the configuration of these parameters that produces the best
performance. Currently, the input parameter values are decided on the basis of prior experience or by the use of brute force techniques. 
A parameter optimization approach that is able to automatically tune these parameters for high performance is greatly desirable. 
In the current work, we explore various machine learning based optimization methods to devise a suitable technique to optimize parameters in the complex 
track reconstruction environment. These methods are evaluated on the basis of a metric that targets high efficiency while keeping
the duplicate and fake rates small. We focus on derivative free optimization approaches that can be applied to problems involving 
non-differentiable loss functions. For our studies, we consider the tracking algorithms defined within A Common Tracking
Software (ACTS) framework. We test our methods using simulated data from ACTS software corresponding to the ACTS Generic detector and the ATLAS ITk detector geometries.
\end{Abstract}

\vfill

\vfill
\end{titlepage}
\def\thefootnote{\fnsymbol{footnote}}
\setcounter{footnote}{0}
%

\normalsize 


\section{Introduction}
\label{intro}
The reconstruction of charged particles' trajectories, commonly known as tracking, is one of the most important but computationally extensive tasks of the full event 
reconstruction chain. 
Precise and efficient measurements of charged particle trajectories are critical for the entire physics program: from reconstructing low momentum tracks in order to 
identify primary vertices, through using medium momentum tracks to identify heavy flavor decays and calculate jet energies, to accurately measuring high momentum messengers 
of electroweak particles or new physics. The underlying tracking software consists of a number of mathematically complex algorithms, each dealing with a particular tracking
sub-process. These algorithms depend on a number of input parameters whose values need to be determined. Great efforts are put into optimizing these input parameters
as these efforts yields both resource savings and physics performance improvements. However, it is  difficult to know the configuration of 
these parameters that can produce the best performance. The values of these input parameters are highly dependent on the underlying tracking geometry and material 
configuration. They are also affected by other factors such as the number of simultaneous pp collisions known as pile-up , center-of-mass energy of collision, target physics process, etc. 
A different set of these parameter values is needed if underlying tracking geometry or material configuration changes. Currently, most of these optimizations are performed using the physicist's previous experience or 
specialized brute-force techniques. Approaches that can automatically tune these parameters for high performance are greatly desirable.  Hyperparameter tuning algorithms used in machine learning offer a number of different approaches in which the optimal parameters can be learned.  These approaches 
have the potential to reduce the effort spent in finding the configurations, as well as they can scan a larger range of parameters, allowing for potentially 
more efficient solutions. With the upcoming High Luminosity Large Hadron Collider (HL-LHC) upgrade, we expect more complex tracking environment which would require
more refined parameter tuning. 

In this study, we have explored five different optimization approaches, which include: Evolutionary Algorithms~\cite{EA_ref}, 
Optuna Hyperparameter optimization~\cite{Optuna_ref}, Lipschitz Optimization~\cite{LIPO_ref}, 
Scikit Optimization~\cite{SciOpt_ref} and Orion Hyperparameter tuning~\cite{Orion_ref}. These approaches are described in more detail later. Initial studies on this topic were presented in~\cite{vchep}.  This work expands on that initial work, include a systematic study of several optimization algorithms and two components of charged particle reconstruction.

\section{Problem Summary}
\label{summary}
\subsection{Testing Framework} \label{Test_framework}
We have used ``A Common Tracking Software (ACTS)''~\cite{ACTS_ref} for our studies which is a community-driven tracking software suite consisting of 
high-level track reconstruction modules. ACTS provides track reconstruction algorithms within a generic, experiment-independent open-source software toolkit.
It includes data structures and algorithms for performing track reconstruction in addition to a tool for fast track simulation. The algorithms are designed to be 
inherently thread-safe to support parallel code execution. The implementation is designed to be fully agnostic to detector technologies, 
and the event processing framework, so that it can be used with different experiments. A number of detector geometries has already been implemented within ACTS. 
Some of these are:
the Belle-II Detector~\cite{Belle_2}, the ATLAS ITk geometry~\cite{altas-itk}, the sPHENIX Detector~\cite{sphenix} and the LDMX Detector~\cite{LDMX}. The underlying tracking algorithms need to be tuned as per the specific detector geometry in order to achieve ultimate physics performance. Therefore, ACTS provides ideal environment for our studies and allows us to tune same parameters for different detector geometries.  

\begin{figure}[!htb]
  \centering
  \includegraphics[width=0.45\textwidth]{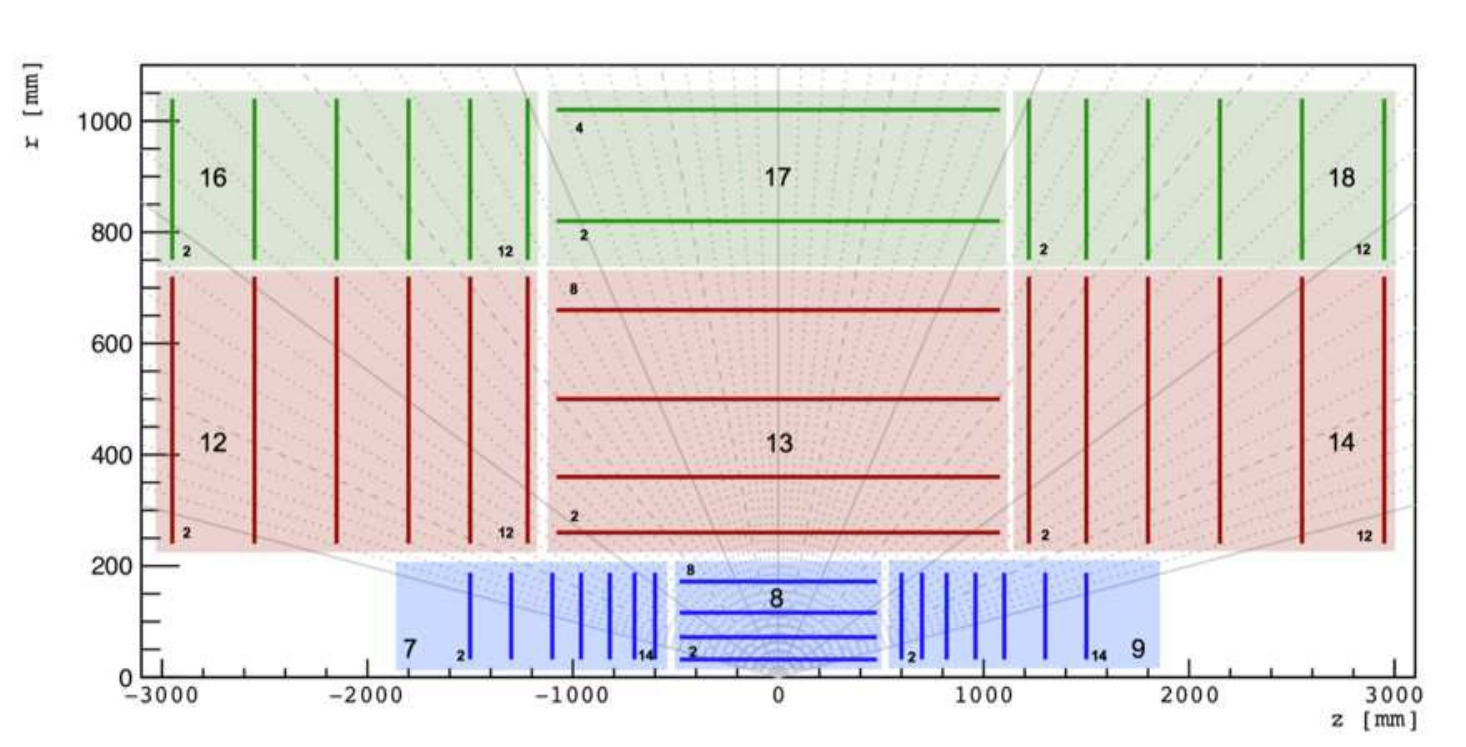}
  \caption{A schematic of ACTS Generic detector used in this study. This figure is taken from ~\cite{ACTS-generic}.}
  \label{fig:genericDet}
\end{figure}

ACTS provides Generic detector geometry which is equivalent to a typical full silicon LHC detector. The Generic Detector is a hermetic detector with silicon based tracker consisting of high granularity pixel layers surrounded by outer tracking layers. A schematic of Generic detector has been shown in Figure~\ref{fig:genericDet}. 
In order to perform our studies, we have used 14TeV $t\bar{t}$ dataset with pile-up = 140 and 200. The data is generated using Pythia8~\cite{pythia8} and simulated through
Generic detector using the ACTS fast simulation algorithm. The input particles are required to have $P_{T}$ $>$ 1GeV and $|\eta|$ $<$ 2.5.     
 
\subsection{Tracking Algorithms}
Two different tracking algorithms from the ACTS tracking framework were considered for these studies:
\begin{itemize}
\item {\bf Track finding/fitting}: The combinatorial kalman filter (CKF)~\cite{ckf} algorithm implemented within ACTS framework is used. 
CKF combines both track finding and fitting in a 
tree-search based algorithm. In ACTS, the main bottleneck in CKF performance is the track seeding algorithm, the algorithm that provides initial track seeds to CKF. There are around 20 parameters in track seeding algorithm. Many of these parameters are dependent on the underlying detector geometry. However, other geometry independent parameters can significantly impact track seeding performance. We have chosen to optimize the following geometry independent  parameters:
\begin{enumerate}
\item {\bf maxPtScattering}: Upper $P_{T}$ limit for scattering angle\footnote{the angle by which particle scatters after colliding with detector material.} calculations.
\item {\bf impactMax}: maximum value for impact parameter.
\item {\bf deltaRMin}: minimum distance in $r$ between two measurements within one seed.
\item {\bf deltaRMax}: maximum distance in $r$ between two measurements within one seed.
\item {\bf sigmaScattering}: number of sigma used for scattering angle calculations.
\item {\bf radLengthPerSeed}: average radiation lengths of material on the length of a seed.
\item {\bf maxSeedsPerSpM}: number of 3-D space-points in top and bottom layers considered for compatibility with middle space-point.
\item {\bf cotThetaMax}: maximum cotTheta angle between two space-points in a seed to be considered compatible.
\end{enumerate}
\item {\bf Vertex finding/fitting}: The Adaptive Multi Vertex Finder (AMVF)~\cite{AMVF} algorithm implemented within ACTS framework is used. AMVF performs simultaneous vertex\footnote{Actual proton-proton collision points.} finding and 
multi-vertex fitting with the help of a multi-vertex fitter. We have considered the following 5 parameters of AMVF for optimization:
\begin{enumerate}
\item {\bf tracksMaxZinterval}: maximum $z$-interval used for adding tracks to multi-vertex fit.
\item {\bf maxVertexChi2}: maximum $\chi^2$ value for tracks to be compatible with fitted vertex.
\item {\bf maxMergeVertexSignificance}: maximum significance on the distance between two vertices to allow merging.
\item {\bf minWeight}: minimum weight assigned to the track for track to be considered compatible with vertex candidate.
\item {\bf maximumVertexContamination}: maximum vertex contamination value.
\end{enumerate}
\end{itemize}

\subsection{Optimization Algorithms Explored}
We have performed optimization of CKF and AMVF parameters using 5 different optimization algorithms. Optimized parameter performance corresponding to each optimization
algorithm has been compared with the default parameter performance of ACTS. 

Each algorithm starts with a random initial parameter configuration chosen from the input 
parameter range. The algorithms run for a number of iterations and, at each iteration, evaluate the score of the current parameter configuration where the score/objective 
function is based on the 
tracking algorithm performance. At each new iteration, the algorithms try to maximize or minimize the score by providing better parameter
configuration using dedicated parameter estimation methods. 
The algorithms use different estimation methods. They typically keep track of previous configurations and 
scores to get better configurations at new iterations. An overview of these optimization algorithm is provided below:
\begin{itemize}
\item {\bf Evolutionary Algorithms (EA)~\cite{EA_ref}}: This algorithm is based on a process analogous to the genetic evolution in living organisms. It starts by initializing
a population of individuals where each individual refers to one configuration of the input parameter set. We considered a population of size 50 with each individual 
assigned to same initial configuration. Then, at each iteration, called a generation, the fitness of each individual is computed using a customized score function based on 
the tracking or vertexing algorithm performance. After the fitness evaluation, simultaneous selections and mutations are performed on  randomly chosen individuals from the population 
such that better performing 
ones are more likely to stay for next generation, and worse performing ones are more likely to be removed. We used a total of 16 generations for the evolution of our parameters.
This algorithm maximizes the score function and provides the configuration with the best score in the end.    
\item {\bf Optuna HyperParameter Optimization (Optuna)~\cite{Optuna_ref}}: Optuna is an optimization framework that consists of multiple hyper-parameter 
sampling algorithms for parameter selection at each trial. For our studies, we have used Tree-structured Parzen Estimator (TPE) which fits one Gaussian Mixture Model 
(GMM) $l(x)$ to the set of parameter values associated with the best objective values, and a second GMM $g(x)$ to the remaining parameter values. It chooses the parameter 
value $x$ that maximizes the quasi-likelihood ratio $\frac{l(x)}{g(x)}$.  
\item {\bf Lipschitz Optimization (LIPO)~\cite{LIPO_ref}}: Lipschitz optimization is based on a simple parameter-free mathematical model for finding the best $x$ that 
maximizes the score function $f(x)$. The key idea is to maintain a piecewise linear upper bound of $f(x)$ and use that to decide which $x$ to evaluate at the next step. It defines 
the upper bound of the function $f(x)$ as $U(x)$ = min$_{i}$ $(f(x_{i}) + k . ||x - x_{i}||_{2})$ such that $U(x)$ $\geq$ $f(x)$ $\forall$ $x$,  $k$ is the Lipschitz constant. 
This algorithm picks new points at 
random and checks if the upper bound at the new point is better than the best point seen so far, and if so, it selects this point as the next step to evaluate.  
\item {\bf Scikit Optimization (Skopt)~\cite{SciOpt_ref}}: This is an optimization framework built inside Python's sklearn library~\cite{sklearn}. It consists of a number of
different parameter optimization techniques. For this study, we have used the forest\_optimize algorithm which performs sequential optimization using decision trees. Skopt 
runs over a number of iterations and minimizes the score function to provide the best parameter configuration.  
\item {\bf Or$\acute{\textrm{\i}}$on HyperParameter Tuning (Orion)~\cite{Orion_ref}}: Or$\acute{\textrm{\i}}$on is an asynchronous framework for 
black-box function optimization. It can be used as a command-line interface as well as python interface. It also implements a number of different optimization methods including
random search and grid search. For our studies, we have used the random search method based on the uniform probability distribution provided to the search parameters. 
\end{itemize} 

\subsection{Performance Evaluation: Score/Objective Function}
The score or objective function is one of the most important component of these optimization studies.
Performance and outcome of the optimization algorithms are highly dependent on the form of score function used. 
The score function is constructed using the performance metrics of the underlying tracking algorithm. Positive weights are given to quantities that we want to 
increase while negative weights are given to quantities that we want to decrease. Likewise, higher weights are given to more important quantities.
\subsubsection{Construction of Score Function for the CKF} 
The important metrics determining the performance of CKF algorithm are:
\begin{itemize}
\item {\bf Track Reconstruction Efficiency}: fraction of generated particles that have created at least 9 measurements on the traversed detectors and are matched with reconstructed tracks with greater than 50\% probability.
\item {\bf Fake Rate}: Fraction of reconstructed tracks that are not associated with any truth particle.
\item {\bf Duplicate rate}: Fraction of  reconstructed tracks associated with same truth generated particle.
\item {\bf Run Time}: Time taken in running the CKF algorithm.
\end{itemize}
Based on these performance metrics, we constructed the following score function for measuring the CKF performance for different parameter configurations:
\begin{equation*}
\textrm{Score/Objective Function} = \textrm{Efficiency} - \left(\textrm{FakeRate} + \frac{\textrm{DuplicateRate}}{\textrm{k}} + \frac{\textrm{RunTime}}{\textrm{k}}\right), \textrm{k} = 7
\end{equation*}
where $k$ is tuned for optimal tradeoff between different quantities.

\subsubsection{Construction of Score Function for the AMVF}
The important metrics determining the performance of AMVF algorithm are:
\begin{itemize}
\item {\bf Eff\_total}: Number of vertices reconstructed by AMVF algorithm out of total detector accepted vertices.
\item {\bf Eff\_clean}: Number of reconstructed vertices associated to a single true generated particle out of all vertices within the detector acceptance.
\item {\bf Eff\_split}: fraction of reconstructed vertices where more than one reconstructed vertices are associated with same truth particle.
\item {\bf Eff\_merge}: fraction of reconstructed vertices where one reconstructed vertex is associated to more than one truth particle.
\item {\bf Eff\_fake}: fraction of reconstructed vertices that are not associated to any truth particle.
\item {\bf Resolution}: 
\begin{equation*}
\frac{{\Delta}{R}}{R} = \sqrt{\sum_{\textrm{vertices}}\frac{((x_{\textrm{reco}} - x_{\textrm{truth}})^{2} + (y_{\textrm{reco}} - y_{\textrm{truth}})^{2} + (z_{\textrm{reco}} - z_{\textrm{truth}})^{2})}{x_{\textrm{truth}}^{2} + y_{\textrm{truth}}^{2} + z_{\textrm{truth}}^{2}}}
\end{equation*}
\end{itemize}
Based on these performance metrics of the AMVF, the following score function has been constructed:

\begin{equation*}
  \textrm{Score/Objective Function} = (\textrm{Eff\_total} + 2\textrm{Eff\_clean}) - (\textrm{Eff\_split} + \textrm{Eff\_merge} + \textrm{Eff\_fake} + \textrm{Resolution})
\end{equation*}

\subsection{Integration of two frameworks}
\begin{figure}[!htb]
  \centering
  \includegraphics[width=0.55\textwidth]{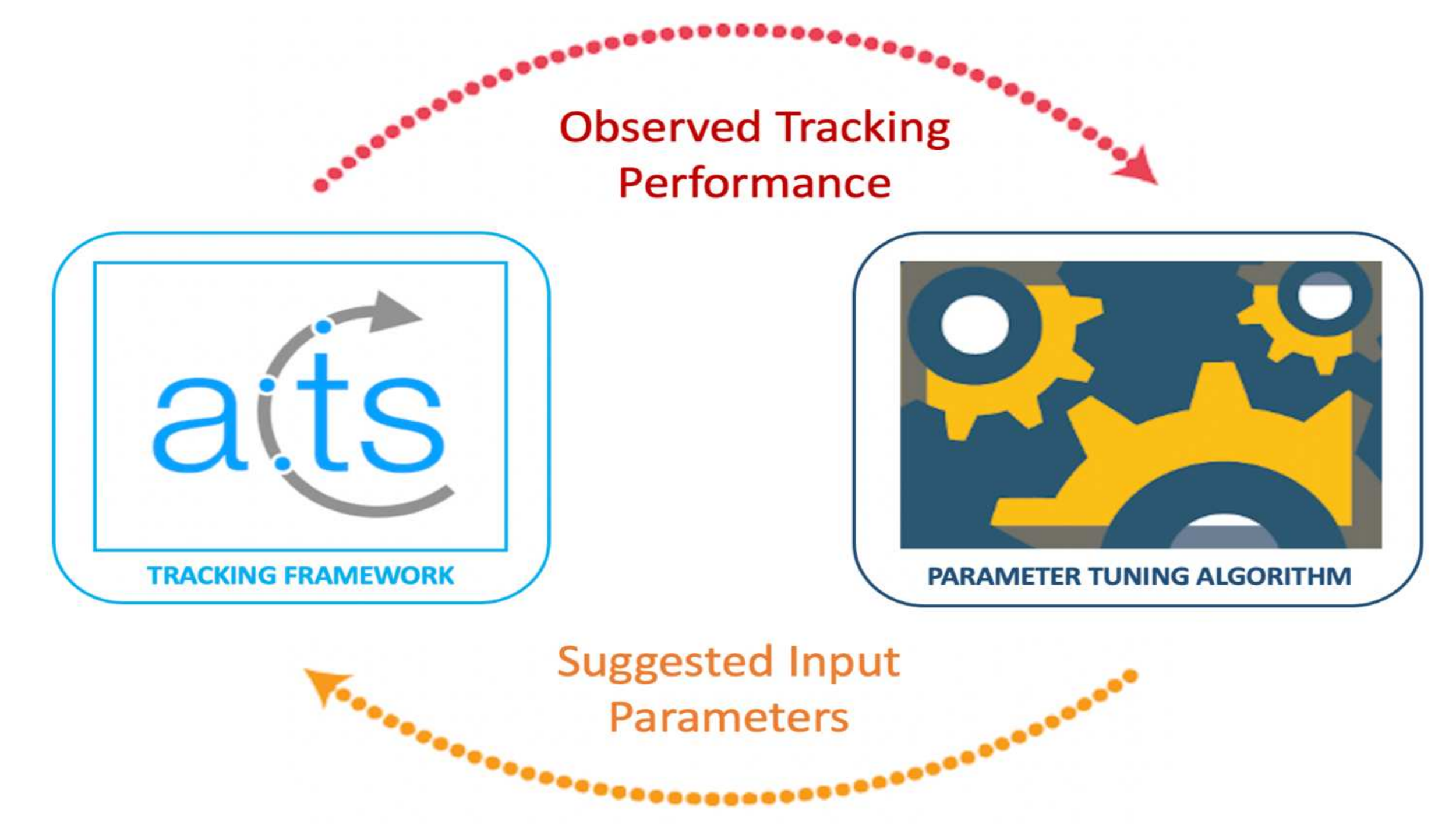}
  \caption{A representation of integration of ACTS tracking software framework with parameter optimization framework.}
  \label{fig:ACTS_Int_Opt}
\end{figure}

In order to perform the optimization studies, we have integrated the ACTS tracking framework with our parameter tuning framework as represented in 
Figure~\ref{fig:ACTS_Int_Opt}. The tuning framework provides input parameter values to the ACTS tracking framework. Next, the tracking algorithm runs with these values and returns the tracking performance to the tuning framework. The tuning framework computes a score based on the tracking performance and runs a parameter estimation method
to evaluate new configuration of input parameters and pass it again to the ACTS framework. This continues for a number of iterations until optimization algorithm achieves
satisfactory score and corresponding parameter configuration with high performance.

\section{Results}
Performance metrics for the CKF and the AMVF have been evaluated corresponding to different optimized parameter configurations and compared with the default parameter configuration.
\subsection{CKF Optimization Results}
\begin{figure}[!htb]
  \centering
  \includegraphics[width=0.39\textwidth]{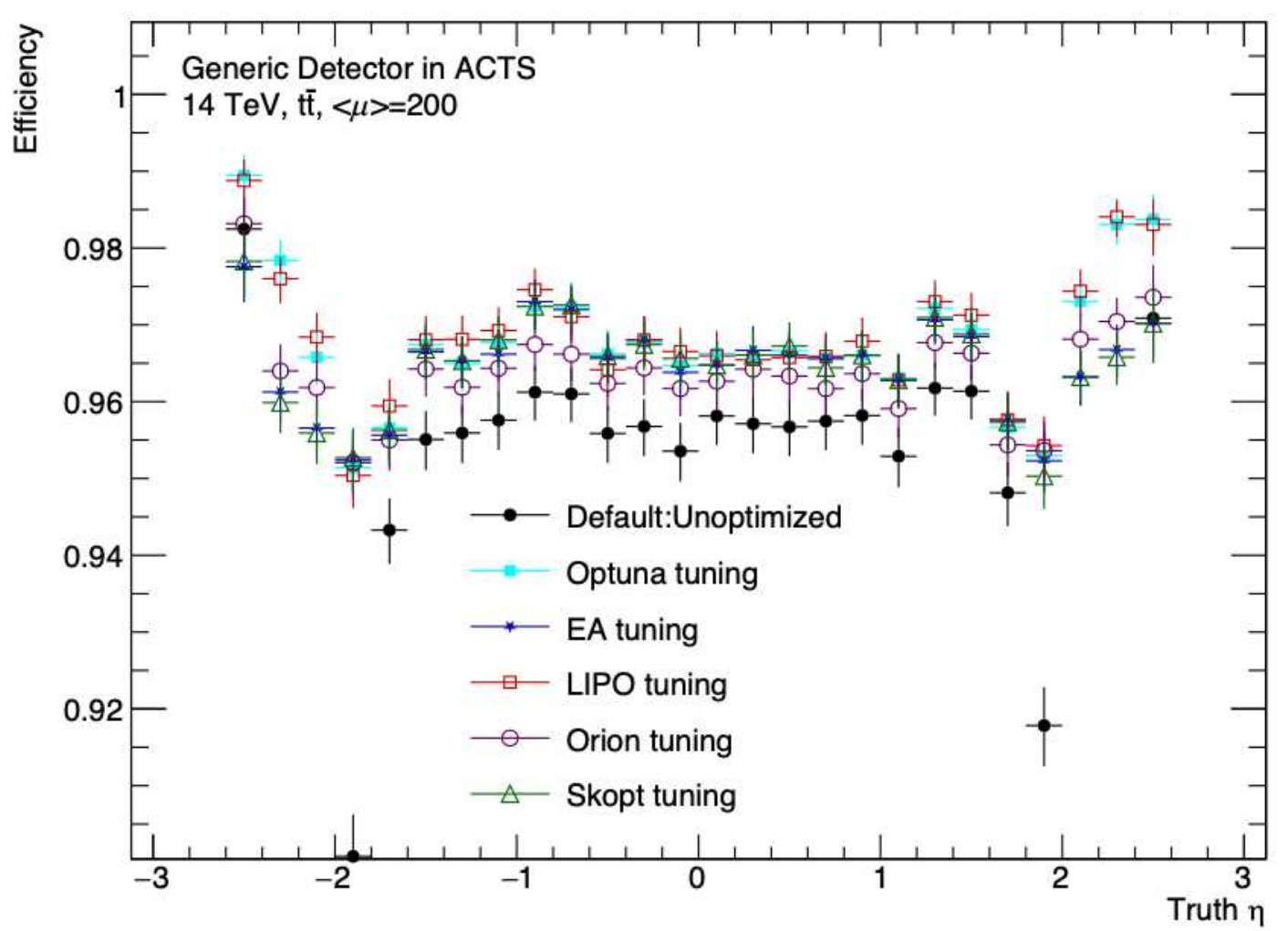}
  \includegraphics[width=0.39\textwidth]{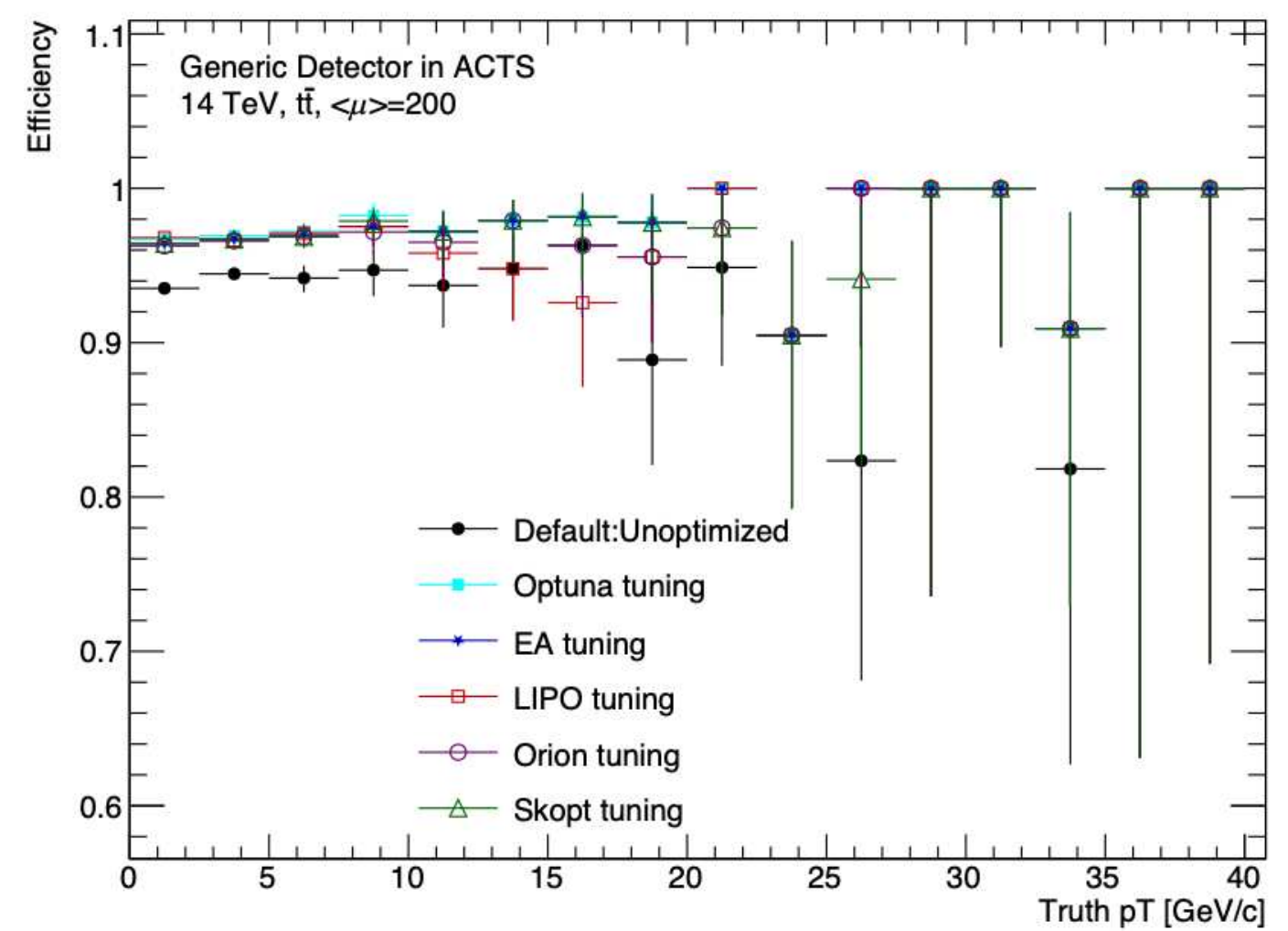} \\
  \includegraphics[width=0.39\textwidth]{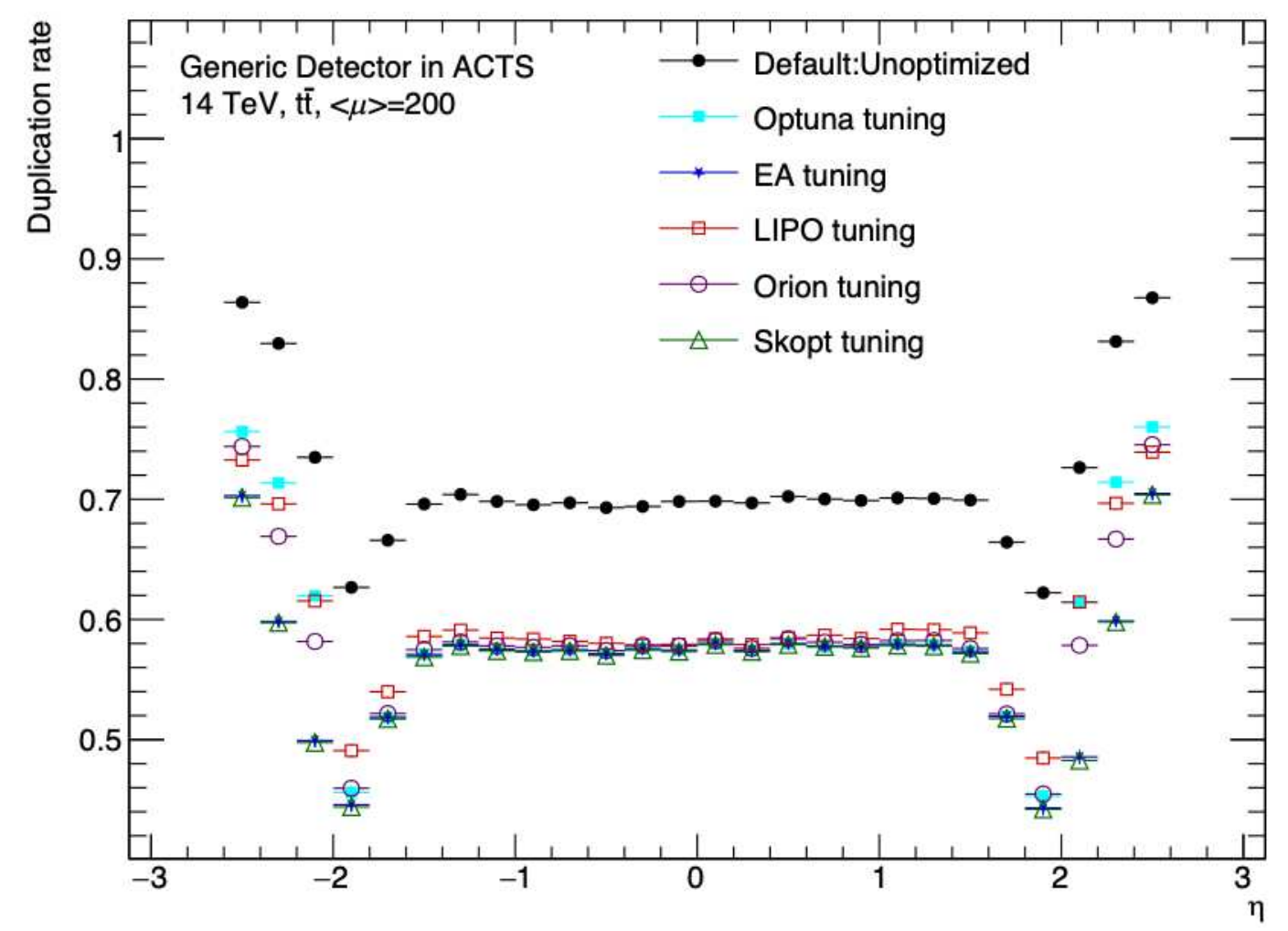}
  \includegraphics[width=0.39\textwidth]{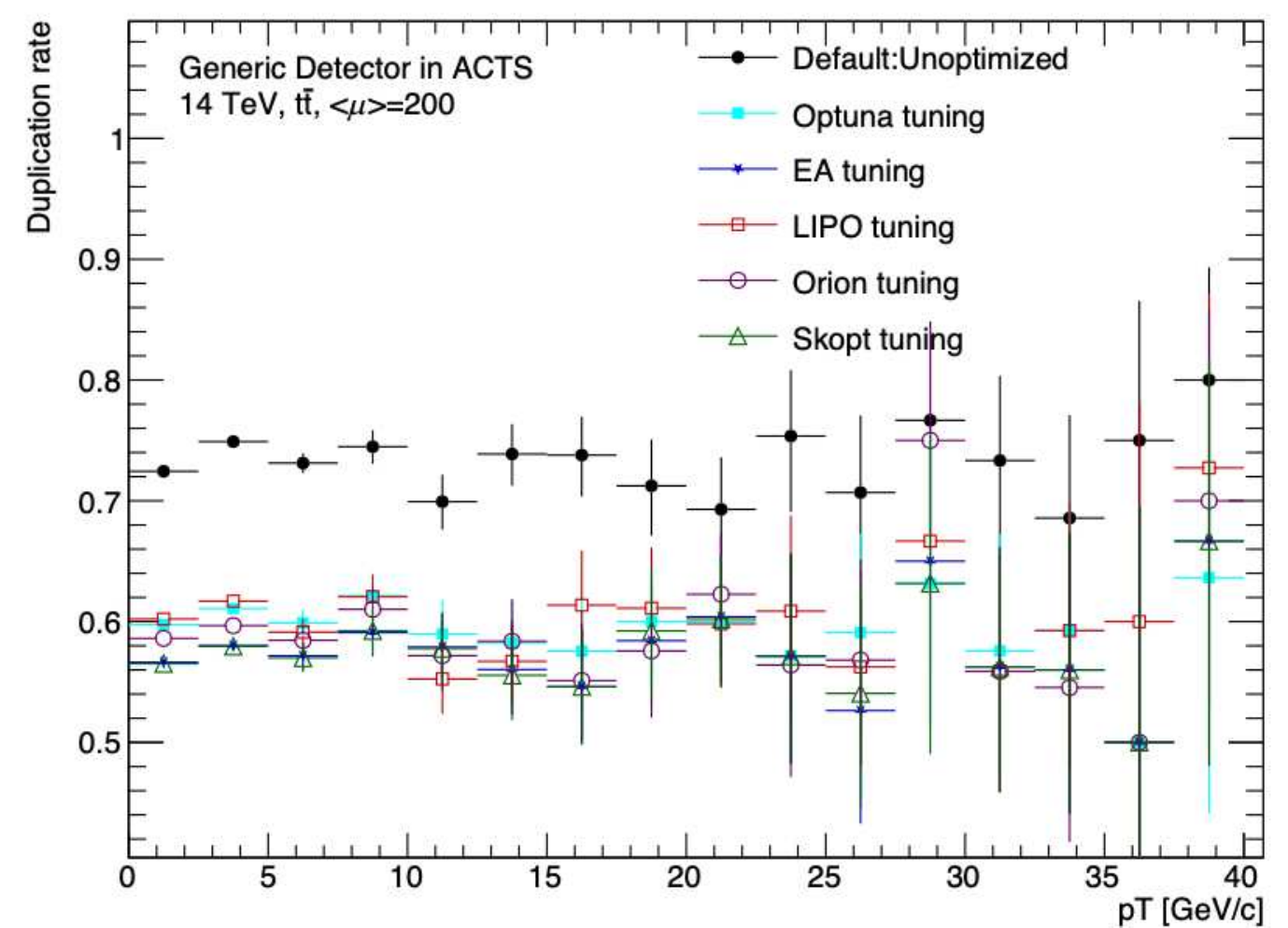}
  \caption{Comparison of CKF performance between different optimized configurations and default configuration for tracking efficiency vs. $\eta$ (upper left), tracking efficiency vs $P_{T}$ (upper right), duplicate rate vs. $\eta$ (lower left) and duplicate rate vs. $P_{T}$ (lower right). Black solid circles represent performance corresponding to default configuration while colored markers represent optimized configurations. }
  \label{fig:CKF_perf_comp_generic}
\end{figure}

The track reconstruction efficiency and duplicate rate as a function of track $P_{T}$ and $\eta$ for different parameter configurations 
is presented in Figure~\ref{fig:CKF_perf_comp_generic}. The black solid circles correspond to the default configuration while the colored markers represent optimized
configurations. A clear improvement in performance is observed
with optimized parameters over all the $P_{T}$ and $\eta$ range, specially in high $\eta$ range.
The table~\ref{tab:CKF_metrics_table} shows the improvements in terms of the average values of the metrics for different configurations.
The fake rate is negligible for both the default and optimized parameters.   

\begin{table}[!htb]
  \begin{center}
    \begin{tabular}{l|c|c|c|c|c|c|}
      \hline
      \hline
       &  Default & EA & Optuna & LIPO & Skopt & Orion \\
      \hline
      Score &  69.06 & 78.88 & 75.38 & 77.28 & 77.05 & 77.79 \\
      Efficiency & 93.6\% & 96.5\% & 96.7\% & 96.8\% & 96.5\% & 96.3\% \\
      Duplicate Rate & 72.6\% & 56.8\% & 59.8\% & 60.3\% & 56.6\% & 58.7\% \\ 
      Fake Rate & 5.56E-03\% & 6.2E-03\% & 5.2E-03\% & 6.8E-03\% & 5.7E-03\% & 8.8E-03\% \\
      time/event (sec) & 50.2 & 31.1 & 46.8 & 37.2 & 40.2 & 33.9 \\
      \hline
      \hline
    \end{tabular}
    \caption{CKF performance metrics for default and optimized configurations.}
    \label{tab:CKF_metrics_table}
  \end{center}
\end{table}

\begin{figure}[!htb]
  \centering
  \includegraphics[width=0.8\textwidth]{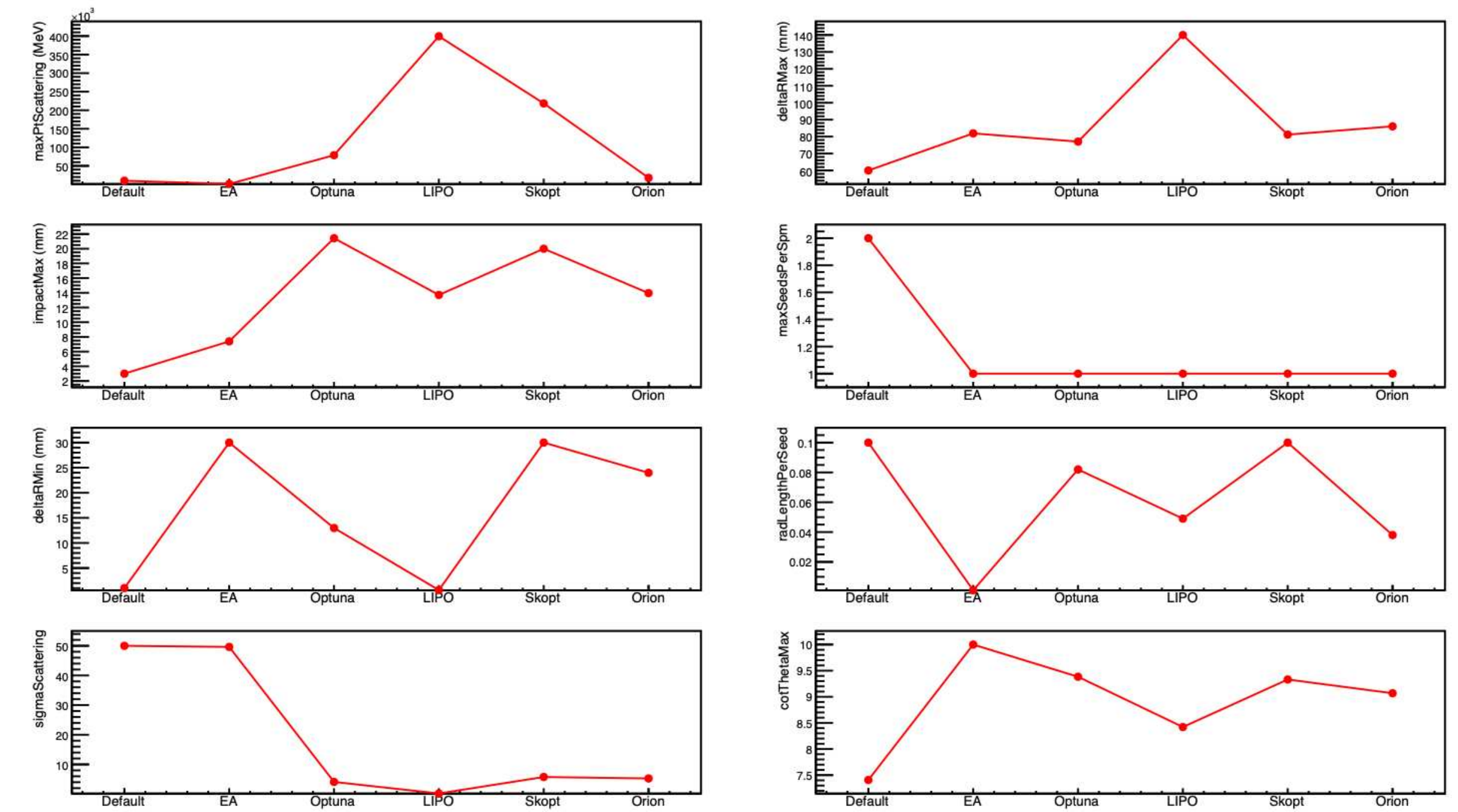}
  \caption{Parameter configurations obtained for CKF from different optimization algorithms.}
  \label{fig:CKF_parVar_generic}
\end{figure}

The parameter values obtained using different optimization algorithms are found to differ from each other as can be seen in Figure~\ref{fig:CKF_parVar_generic}. This
variation clearly states that there are multiple local minima in the parameter space of CKF performance. 

\subsection{AMVF Optimization Results}
\begin{figure}[!htb]
  \centering
  \includegraphics[width=0.32\textwidth]{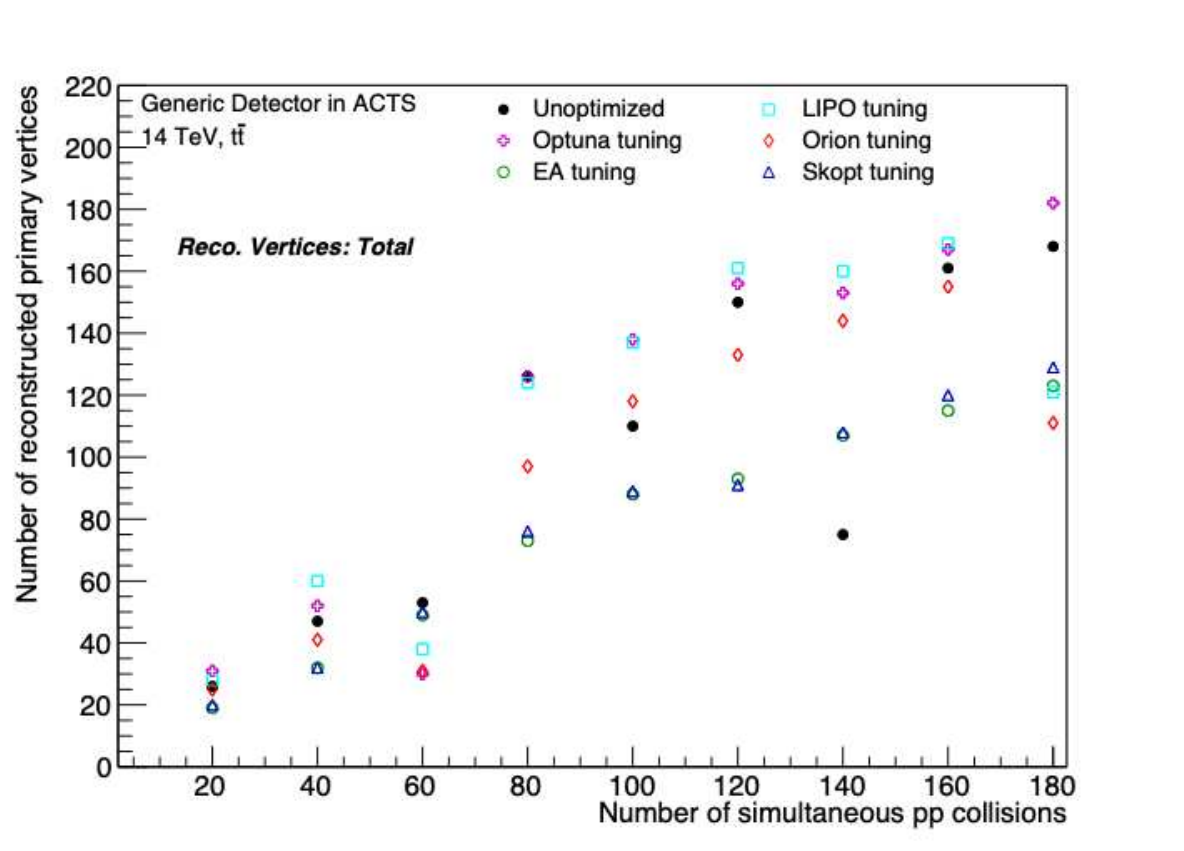}
  \includegraphics[width=0.32\textwidth]{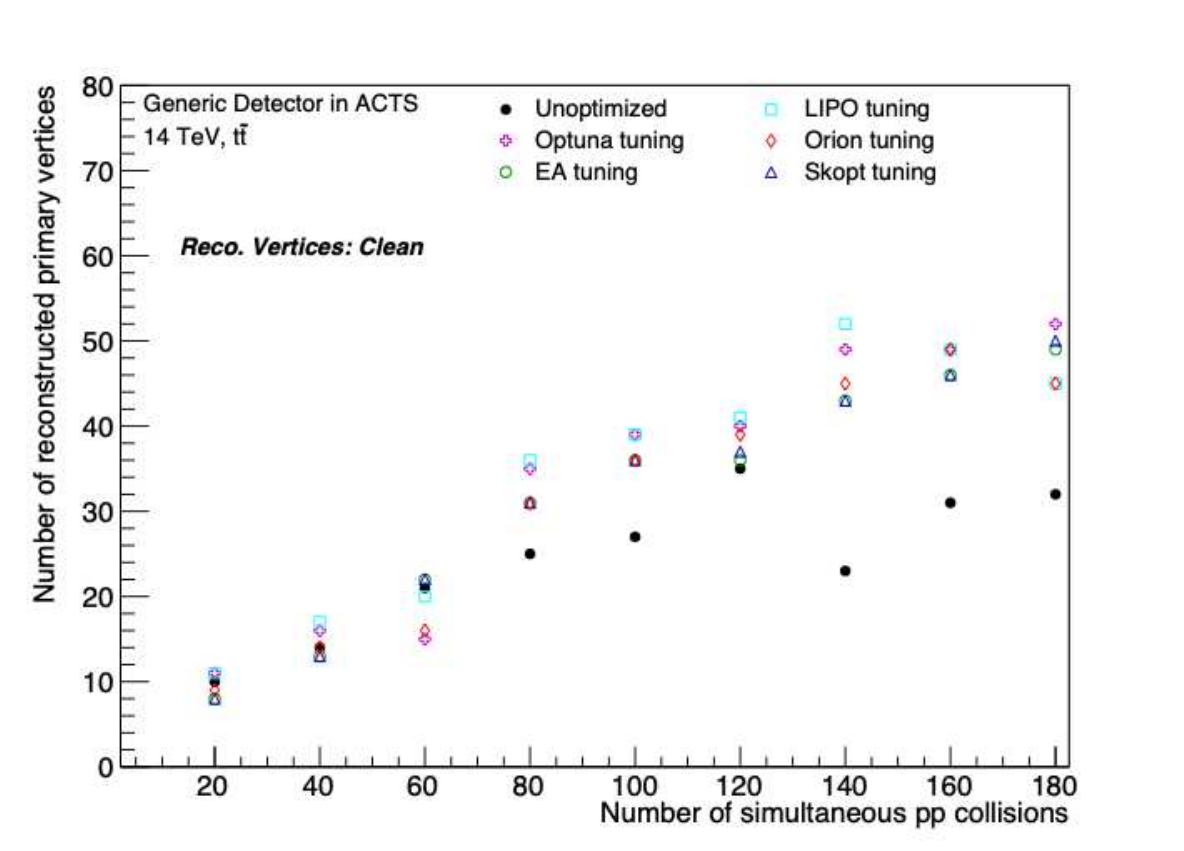} 
  \includegraphics[width=0.32\textwidth]{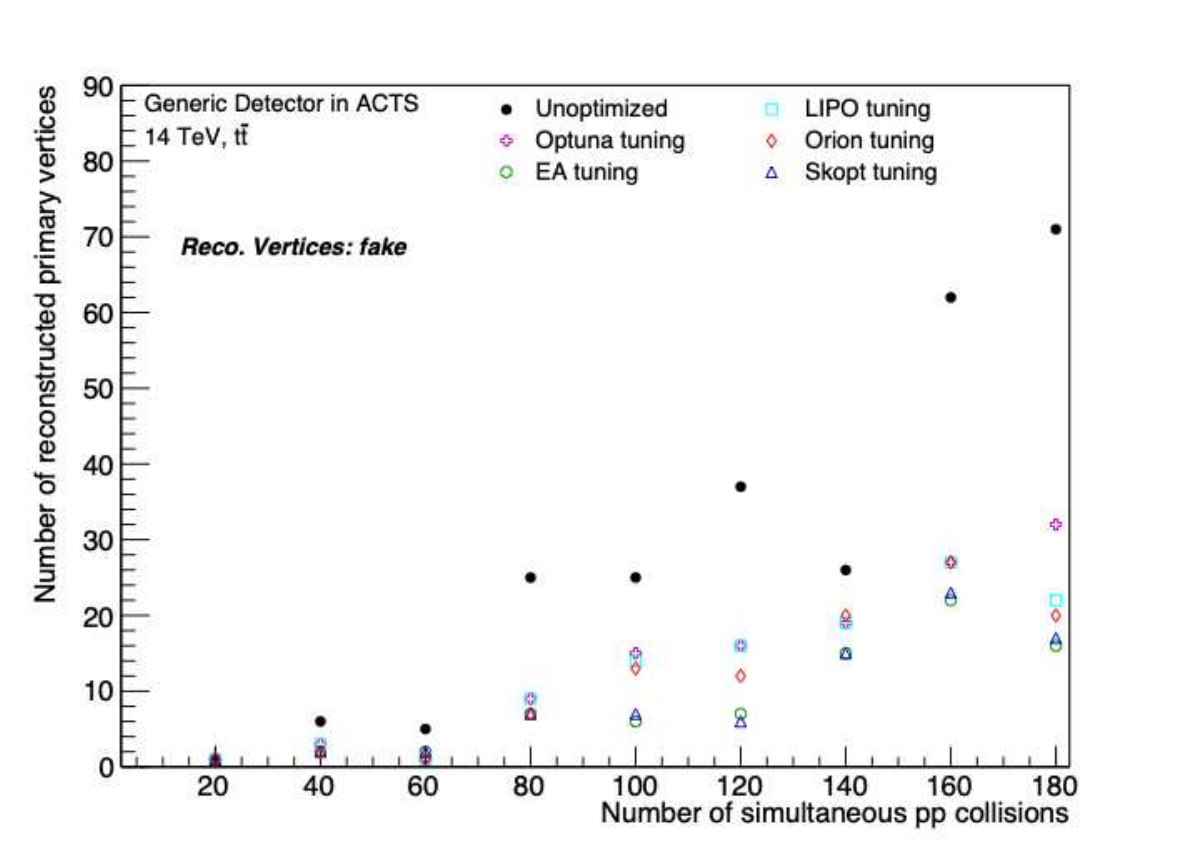} \\
  \includegraphics[width=0.32\textwidth]{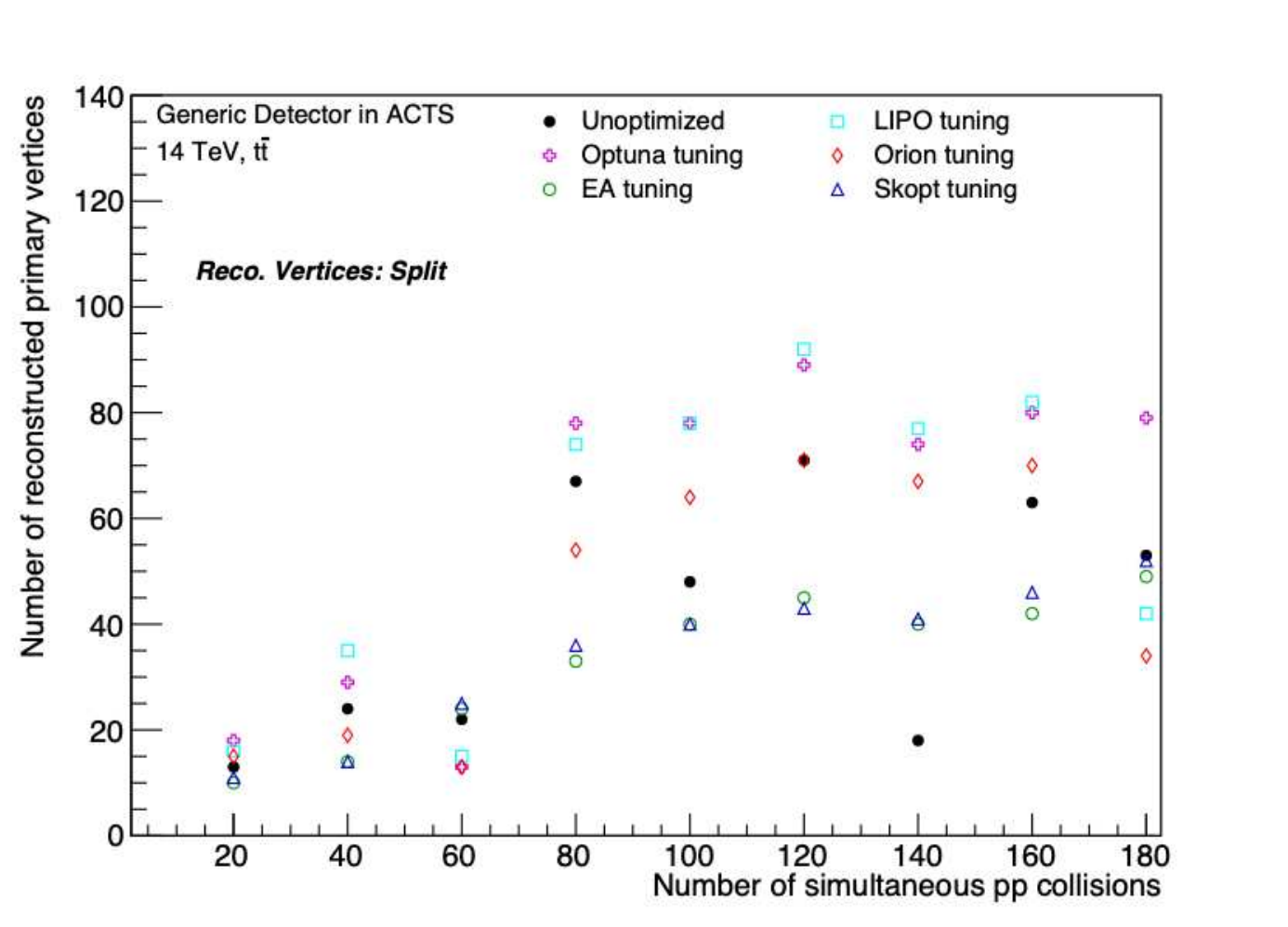}
  \includegraphics[width=0.32\textwidth]{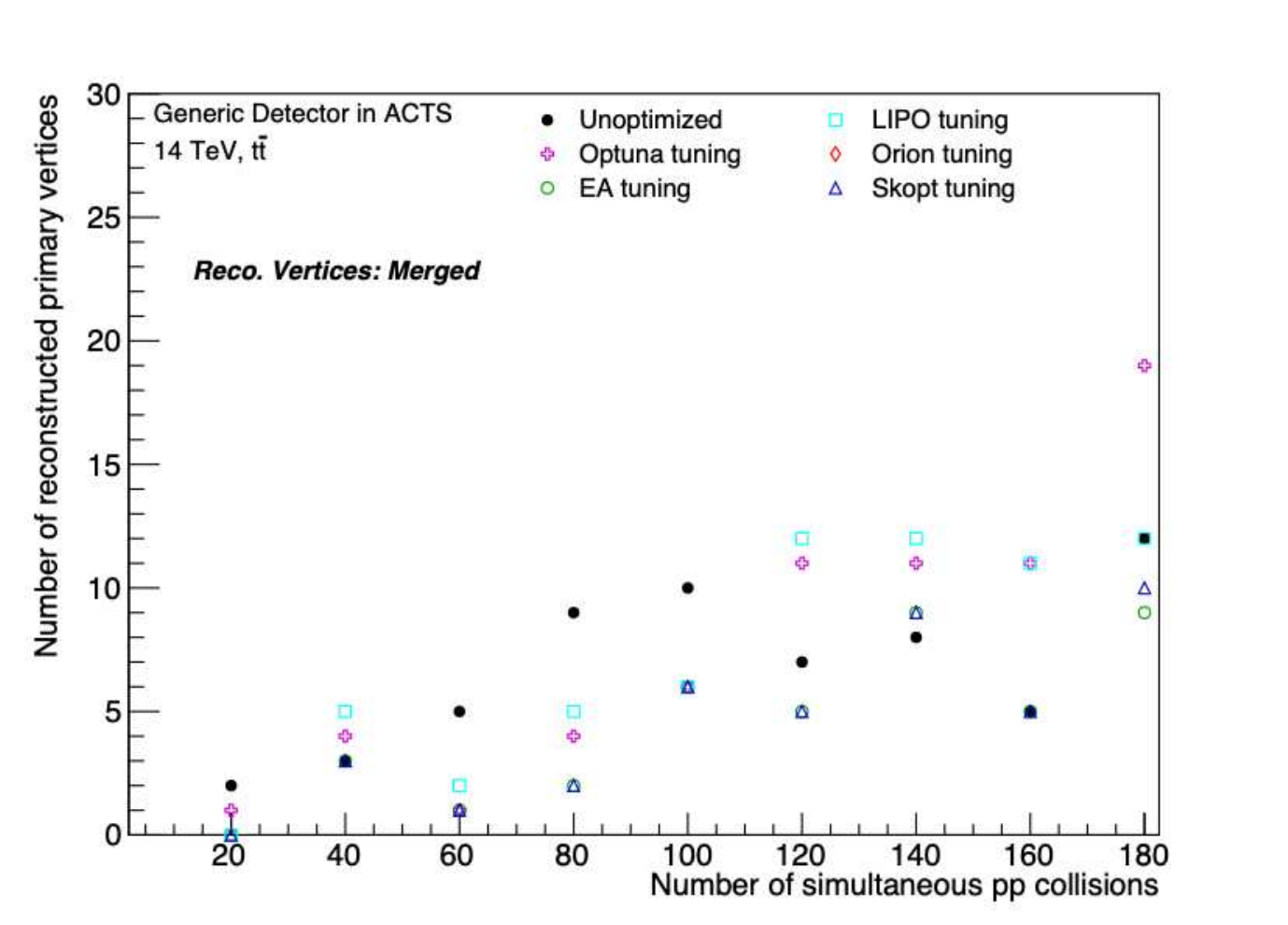}
  \caption{Comparison of AMVF performance between different optimized and default configurations for total (upper left), clean (upper middle), fake (upper right), split (lower left) and merged (lower right) reconstructed vertices. Black solid circles represent performance corresponding to the default configuration while colored ones represent performance corresponding to the optimized configurations. }
  \label{fig:AMVF_perf_comp_generic}
\end{figure}

\begin{figure}[!htb]
  \centering
  \includegraphics[width=0.32\textwidth]{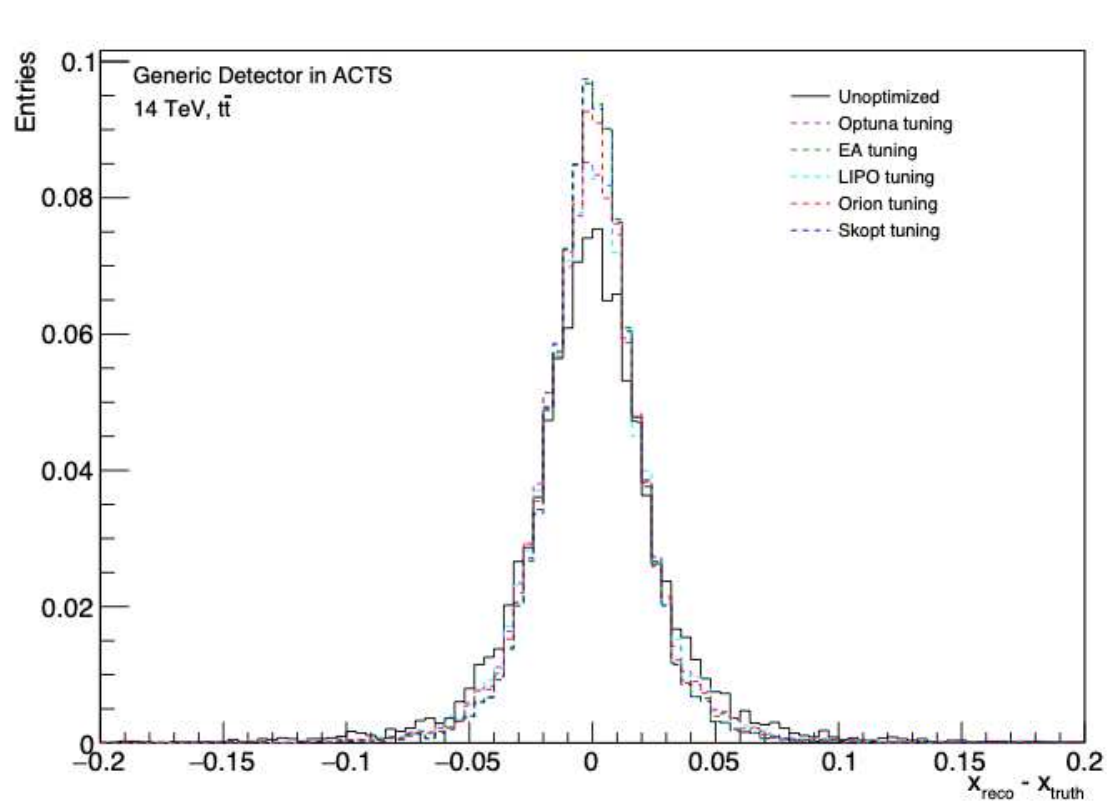}
  \includegraphics[width=0.32\textwidth]{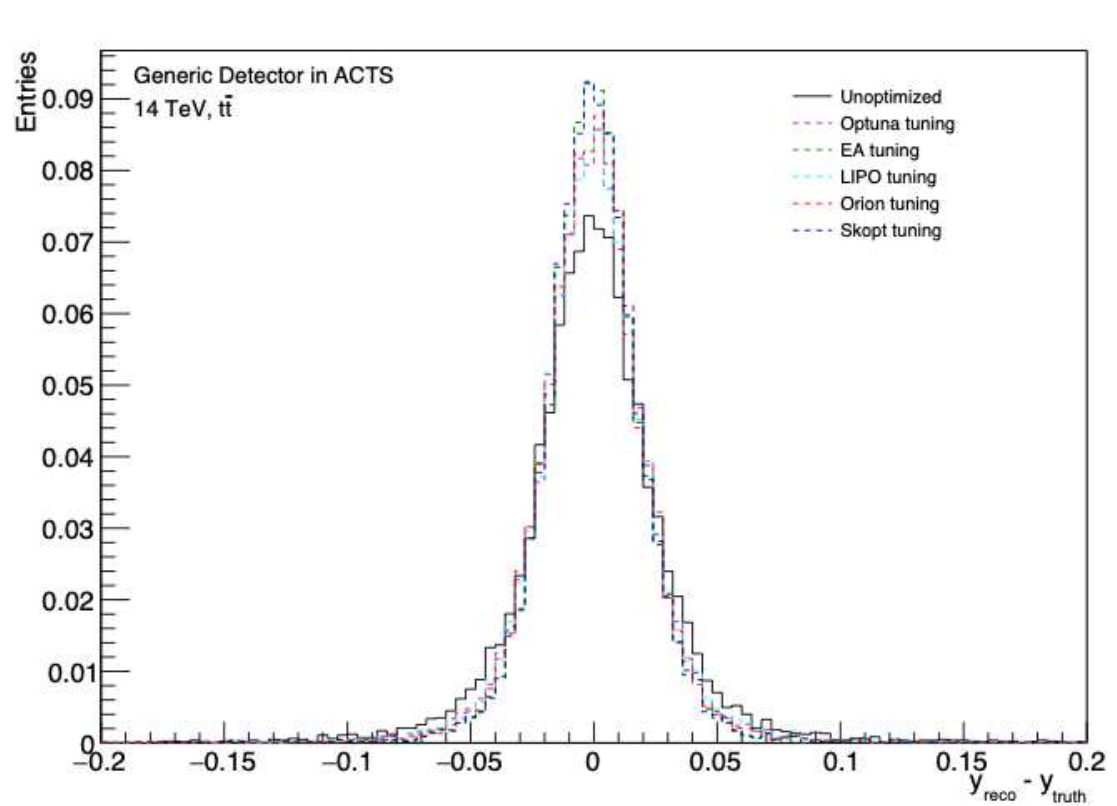} 
  \includegraphics[width=0.32\textwidth]{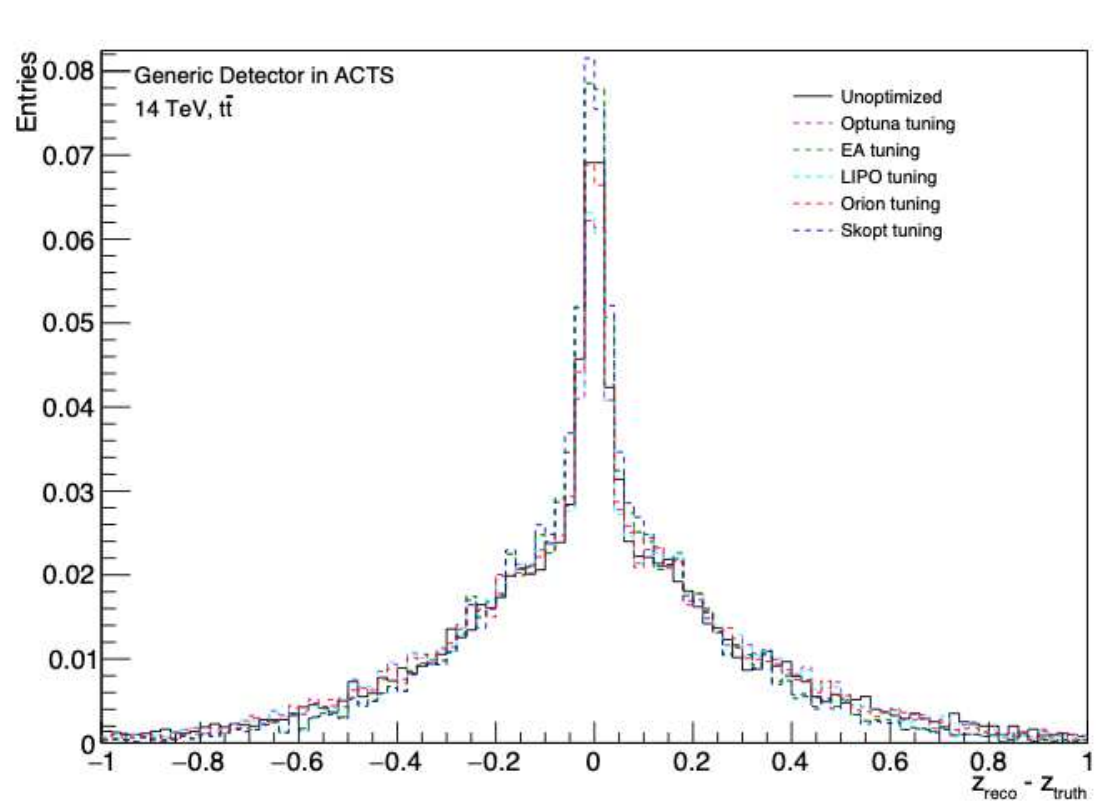} 
  \caption{Comparison of AMVF resolution in $x$, $y$ and $z$ for different optimized configurations and default configuration. Black lines represent resolution corresponding to the default configuration while colored lines represent the optimized configurations. }
  \label{fig:AMVF_resolution_comp_generic}
\end{figure}

\begin{figure}[!htb]
  \centering
  \includegraphics[width=0.7\textwidth]{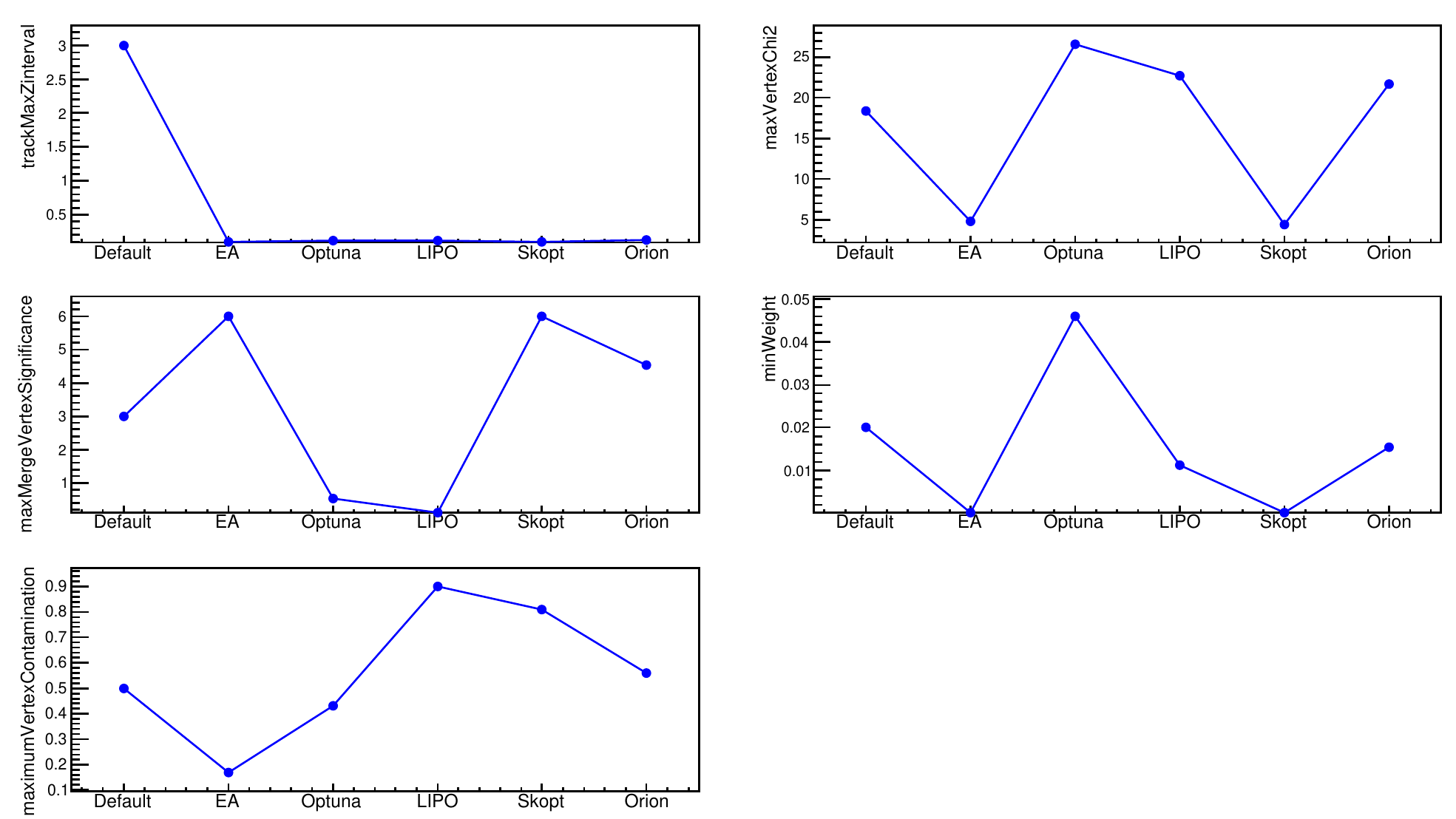}
  \caption{Parameter configurations obtained for the AMVF from the different optimization algorithms.}
  \label{fig:AMVF_parVar_generic}
\end{figure}

The number of reconstructed vertices tagged as total, clean, fake, split and merged corresponding to optimized parameters are compared with the default configuration in 
Figure~\ref{fig:AMVF_perf_comp_generic}. With the optimized parameters there are more clean and fewer fake vertices compared to the default. However, there is a  trade-off 
between the total and split vertices. The position resolution in $x$, $y$ and $z$ are compared in
Figure~\ref{fig:AMVF_resolution_comp_generic} and shows more centered $z$-resolution for optimized configurations. The parameter values obtained by different 
optimization algorithms are presented in Figure~\ref{fig:AMVF_parVar_generic}, again emphasizing on the presence of multiple local minima in AMVF space.

\section{CKF optimization using the ATLAS ITk geometry within ACTS}
\begin{figure}[!htb]
  \centering
  \includegraphics[width=0.4\textwidth]{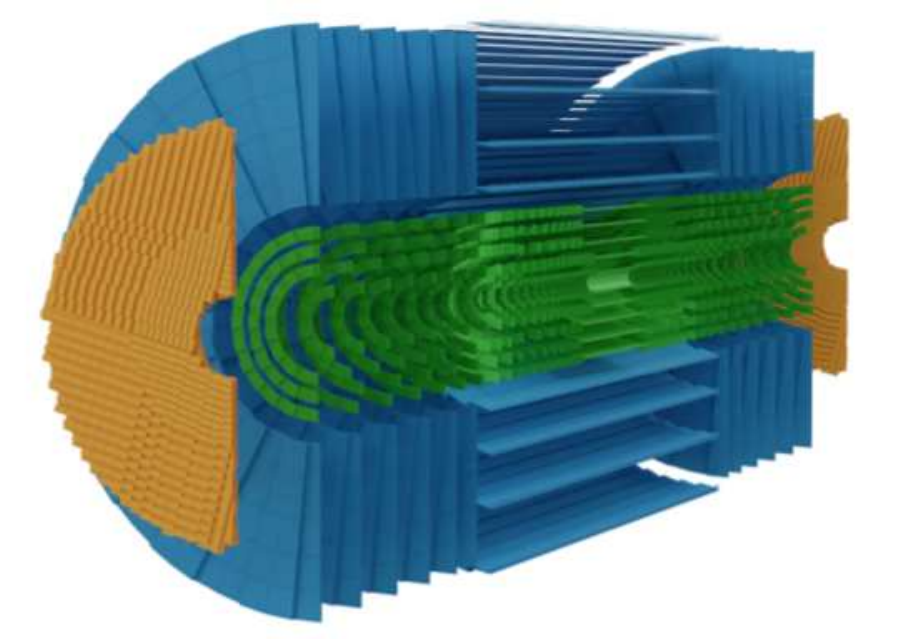}
  \includegraphics[width=0.4\textwidth]{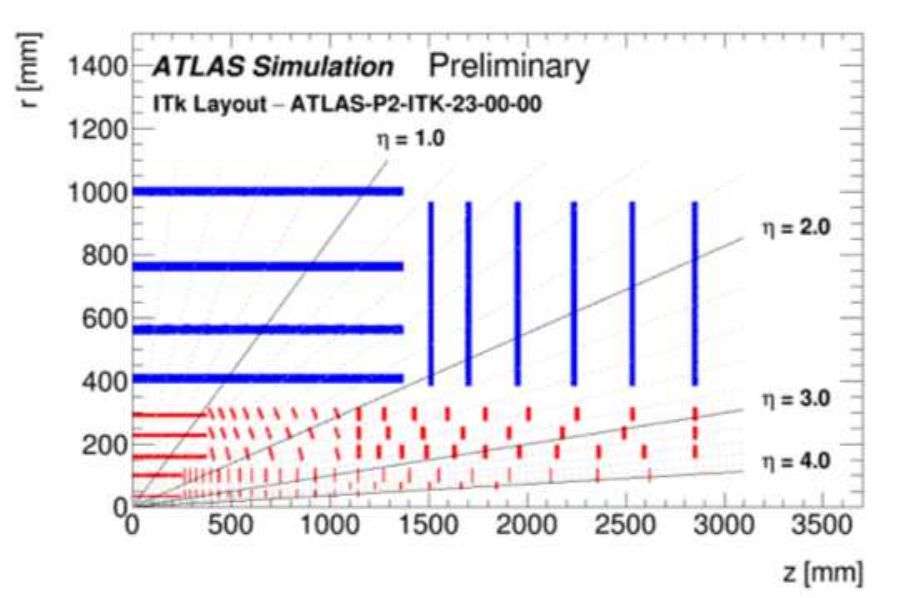}
  \caption{Diagram showing ATLAS ITk geometry.}
  \label{fig:ATLAS-ITk_geo}
\end{figure}

We used the same optimization techniques for the ATLAS ITk geometry~\cite{atlas-itk}, which is shown in Figure~\ref{fig:ATLAS-ITk_geo}, within the ACTS framework. 
We optimized the same CKF parameters using $t\bar{t}$ data with pile-up = 200 having $P_{T}$ $>$ 1GeV and $|\eta|$ $<$ 4.0. We used the same CKF score function with $k$ = 5.

The comparison of performance for track reconstruction efficiency and duplicate rate as a function of $P_{T}$ and $\eta$ is shown at~\cite{altas-itk_results}. 




\section{Conclusions}
We have presented the implementation and performance of five different derivative-free optimization algorithms within the context of ACTS software framework and demonstrated our results using ACTS Generic detector and ATLAS ITk detector, both integrated within ACTS framework. We have performed
our studies using Combinatorial Kalman Filter and Adaptive Multi Vertex Finder tracking algorithms. The performance of optimized parameters have been compared with default
configuration currently present within ACTS framework. Our results show that the different optimization algorithms are able to automatically find better 
performing parameter configurations. These studies demonstrate the potential of such algorithms to be used for the automatic tuning of tracking algorithms to different detector geometries.

\Acknowledgements
This work was supported in part by the National Science Foundation under Cooperative Agreement OAC-1836650.


\end{document}